\documentclass[12pt]{elsarticle}
\usepackage{bbm}
\usepackage{amsmath}
\usepackage{amsfonts,amsbsy}
\usepackage{amssymb}
\usepackage{graphics}

\def\empile#1\over#2{\mathrel{\mathop{\kern 0pt#1}\limits_{#2}}}
\def\bs{\boldsymbol}

\newcommand{\slv}{\raise.15ex\hbox{$/$}\kern-.53em\hbox{$v$}}
\newcommand{\slF}{\raise.15ex\hbox{$/$}\kern-.53em\hbox{$F$}}
\newcommand{\slL}{\raise.15ex\hbox{$/$}\kern-.53em\hbox{$L$}}
\newcommand{\slP}{\raise.15ex\hbox{$/$}\kern-.53em\hbox{$P$}}
\newcommand{\slp}{\raise.15ex\hbox{$/$}\kern-.53em\hbox{$p$}}
\newcommand{\slq}{\raise.15ex\hbox{$/$}\kern-.53em\hbox{$q$}}
\newcommand{\slR}{\raise.15ex\hbox{$/$}\kern-.53em\hbox{$R$}}
\newcommand{\slQ}{\raise.15ex\hbox{$/$}\kern-.53em\hbox{$Q$}}
\newcommand{\slK}{\raise.15ex\hbox{$/$}\kern-.53em\hbox{$K$}}
\newcommand{\slk}{\raise.15ex\hbox{$/$}\kern-.53em\hbox{$k$}}
\newcommand{\slD}{\raise.15ex\hbox{$/$}\kern-.73em\hbox{$D$}}
\newcommand{\slC}{\raise.15ex\hbox{$/$}\kern-.53em\hbox{$C$}}
\newcommand{\slA}{\raise.15ex\hbox{$/$}\kern-.53em\hbox{$A$}}
\newcommand{\slSigma}{\raise.15ex\hbox{$/$}\kern-.53em\hbox{$\Sigma$}}
\newcommand{\slpartial}{\raise.15ex\hbox{$/$}\kern-.53em\hbox{$\partial$}}
\newcommand{\slcalP}{\raise.15ex\hbox{$/$}\kern-.63em\hbox{$\cal P$}}

\newcommand\colora{}
\newcommand\colorb{}
\newcommand\colorc{}
\newcommand\colord{}

\renewcommand\vec{}

\def\x{{\boldsymbol x}}
\def\y{{\boldsymbol y}}

\def\E{{\boldsymbol E}}
\def\B{{\boldsymbol B}}

\journal{Nuclear Physics A}

\begin{document}

\begin{frontmatter}



\title{Color Glass Condensate and Glasma}


\author{Fran\c cois Gelis}

\address{Institut de Physique Th\'eorique, CEA/Saclay\\91191, Gif sur Yvette cedex}

\begin{abstract}
  In this talk, I review the Color Glass Condensate theory
  of gluon saturation, and its application to the
  early stages of heavy ion collisions.
\end{abstract}

\begin{keyword}
QCD \sep Gluon saturation \sep Heavy ion collisions


\end{keyword}

\end{frontmatter}

% \linenumbers

\section{Color Glass Condensate}
\label{sec:cgc}
In heavy ion collisions, the vast majority of the produced particles
have a transverse momentum of at most a couple of GeV's. This implies
that they are produced from incoming partons that carry a very small
fraction $x$ of the longitudinal momentum of the projectiles.
Furthermore, this momentum fraction decreases with the center of mass
energy of the collision. At RHIC energy, the typical value of $x$ is
of the order of $10^{-2}$ for the bulk of particle production, and of
order $x\sim 4.10^{-4}$ at the LHC. However, at such small values of
$x$, the gluon density in a proton or nucleus becomes large which
leads to a new phenomenon known as {\sl gluon saturation}, due to
non-linear recombinations among the gluons.
\begin{figure}[htb]
\begin{center}
\resizebox*{!}{3.5cm}{\includegraphics{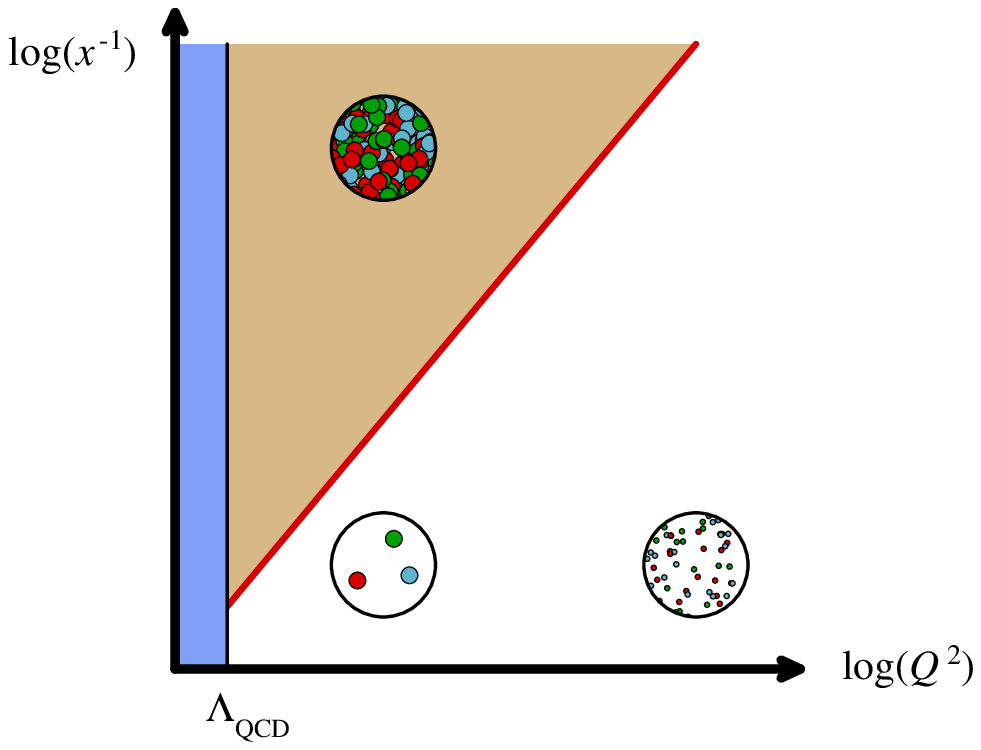}}
\hfil
\resizebox*{!}{3.5cm}{\includegraphics{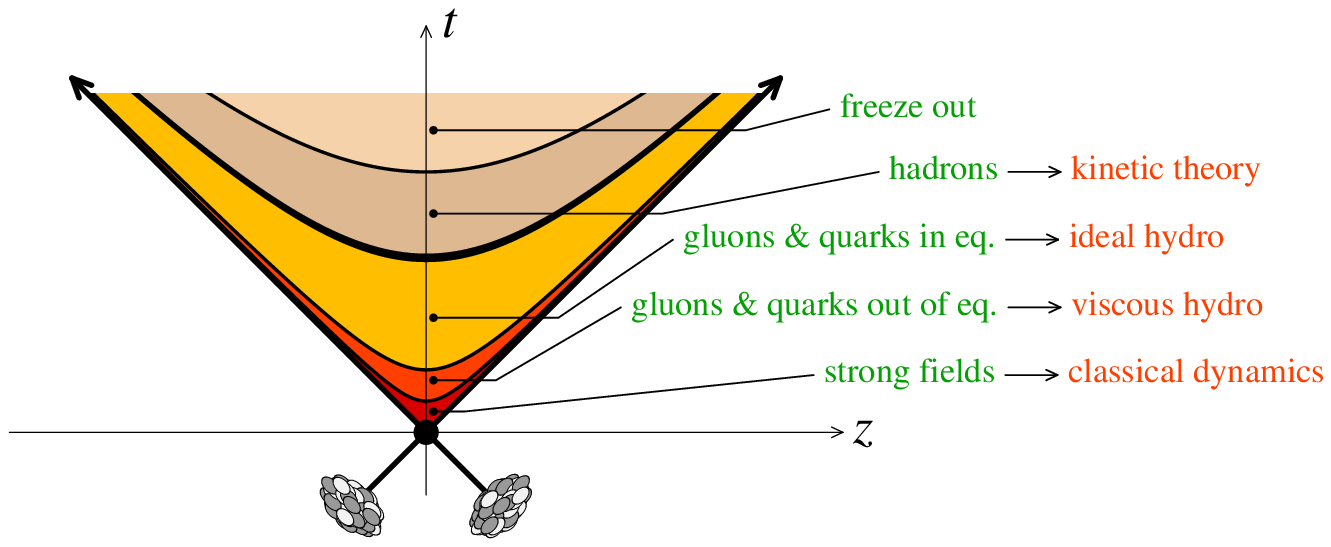}}
\end{center}
\caption{\label{fig:satdomain}Left: Saturation domain in the $x,Q^2$
  plane. Right: Stages of a heavy ion collision.}
\end{figure}
Gluon saturation is characterized by an $x$-dependent momentum scale,
the saturation momentum $Q_s(x)$, that delimitates the domain in
momentum where saturation takes place. This domain is represented in
the figure \ref{fig:satdomain}.

\begin{figure}[htb]
\begin{center}
\resizebox*{!}{3cm}{\includegraphics{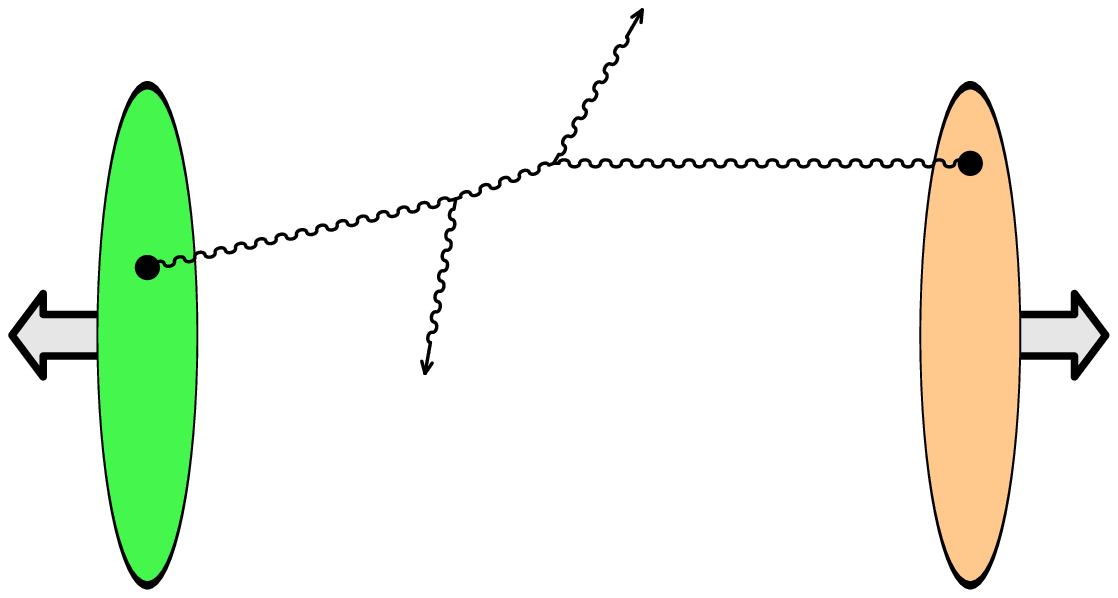}}
\hfil
\resizebox*{!}{3cm}{\includegraphics{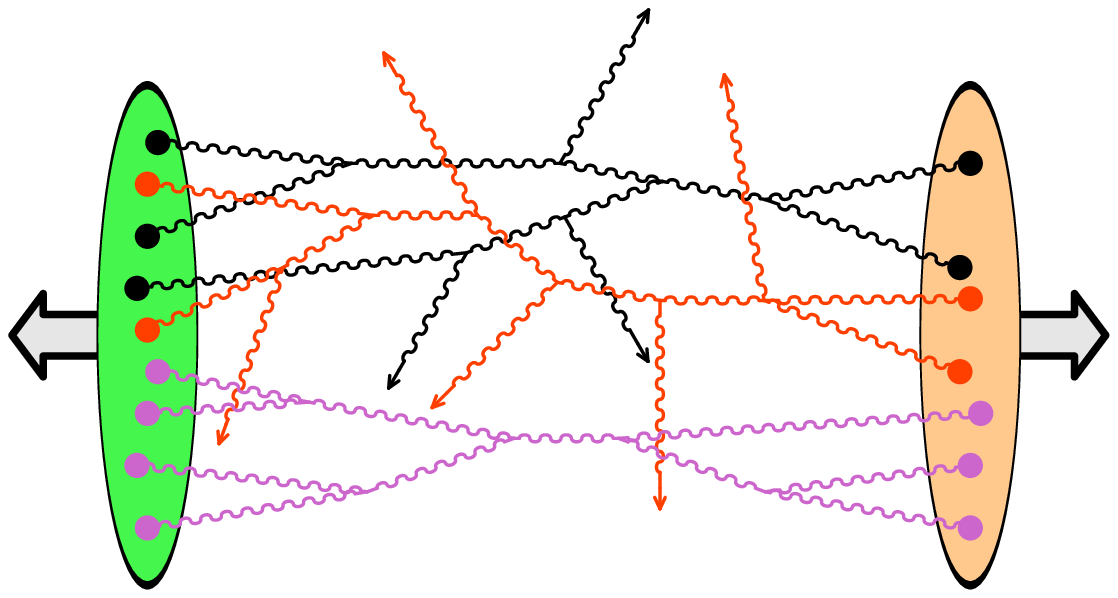}}
\end{center}
\caption{\label{fig:dildense}Left: Gluon production in the dilute regime. Right: Saturated regime.}
\end{figure}

Having identified this region where nonlinear gluon interactions
become important, one faces now the question of how to compute
physical observables reliably in this regime where one collides
projectiles made of a very large number of constituents. The main
novelty in the saturated regime, compared to the dilute one, is that
the typical particle production processes involve multiple gluon
interactions, as shown in the figure \ref{fig:dildense}.
The standard perturbative techniques based on Feynman diagrams are
well suited to handle the dilute situation, but they become
impractical in the saturated regime because an infinite number of
graphs would contribute at each order in $g^2$. In addition, in order
to compute observables in this dense regime, we also need to know the
probability to find these multigluon states in the wavefunctions of
the incoming nuclei.

\begin{figure}[htb]
\begin{center}
\resizebox*{!}{1cm}{\includegraphics{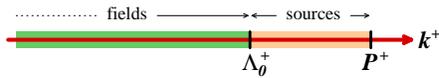}}
\end{center}
\caption{\label{fig:sep}Separation of the degrees of freedom in the CGC.}
\end{figure}

The Color Glass Condensate is an effective description of QCD in the
saturation regime. In this effective theory, one divides the degrees
of freedom (see the figure \ref{fig:sep}) into fast partons with
longitudinal momentum $k^+>\Lambda^+$, and slow partons with
$k^+<\Lambda^+$ that have a significant dynamical evolution over the
time-scales of the reaction of interest
\cite{MV}.
Thanks to Lorentz time dilation, the fast partons are essentially
frozen during the collision process, and it is sufficient to know what
color current they carry.  Therefore, their description in the CGC is
simplified into a distribution of static color currents
$J^\mu=\delta^{\mu+}\rho(x^-,\x_\perp)$ (this is for a projectile
moving in the $+z$ direction). On the contrary, this approximation
does not work for the slow degrees of freedom, and therefore we must
keep describing them as conventional gauge fields $A^\mu$. Due to the
separation in longitudinal momentum between these two types of degrees
of freedom, their coupling is eikonal, $A_\mu J^\mu$. Thus, the
effective Lagrangean of the CGC is:
\begin{equation*}
{\cal L}=\underbrace{-\frac{1}{2}\;{\rm tr}\,F_{\mu\nu}F^{\mu\nu}}_{\mbox{gluon dynamics}}
+\underbrace{\vphantom{\frac{1}{2}}({\colord J^\mu_1+J^\mu_2})}_{\mbox{fast partons}} {\colorb A_\mu}
\; .
\end{equation*}
This description would not be complete without a way to specify the
color charge density $\rho(x^-,\x_\perp)$. This distribution depends
on the precise configuration of the fast partons (e.g. their
localization in the transverse plane at the time of the collision),
which is not known. Thus, it should be considered as a stochastic
quantity, with a probability distribution $W_{\Lambda^+}[\rho]$. The
distribution $W_{\Lambda^+}[\rho]$ plays in the CGC the same role as
partons distributions in conventional low density perturbative QCD.

The CGC effective theory introduces a cutoff $\Lambda^+$ that
separates the fast and the slow degrees of freedom. However, this
cutoff is unphysical and one should ensure that physical observables
do not depend upon it. It turns out that for this to be true, the
probability distribution $W_{\Lambda^+}[\rho]$ must evolve with
$\Lambda^+$ according to the JIMWLK equation \cite{JIMWLK},
\begin{equation*}
\frac{\partial {\colorb W_{\Lambda^+}}}{\partial\ln(\Lambda^+)}
=
{\cal H}\;\;
{\colorb W_{\Lambda^+}}\; ,\quad
  {\cal H}=\frac{1}{2}
  \int\limits_{\vec\x_\perp,\vec\y_\perp}
  \frac{\delta}{\delta{\colord{\alpha}(\vec\y_\perp)}}
  {\colorb\eta(\vec\x_\perp,\vec\y_\perp)}
  \frac{\delta}{\delta{\colord{\alpha}(\vec\x_\perp)}}\; ,
\end{equation*}
where
$-{\bs\partial}_\perp^2\,{\colord{\alpha}(\vec\x_\perp)}=\rho(1/\Lambda^+,\vec\x_\perp)$.
Effectively, the JIMWLK equation resums to all orders the powers of
$\alpha_s\ln(\Lambda^+)$ (or, equivalently, powers of
$\alpha_s\ln(1/x)$), and its kernel $\eta(\x_\perp,\y_\perp)$ includes
all corrections in the color source $\rho$. When approximated to low
density by expanding the kernel to lowest order $\rho$, one recovers
the BFKL equation that describes the evolution to low $x$ of
non-integrated gluon distribution of a dilute hadron.

\section{Just before the collision: factorization}
\label{sec:fact}
A high energy heavy ion collision can be divided in several successive
stages, as shown in the figure \ref{fig:satdomain}. The CGC applies to
the description of the wavefunctions of the incoming projectiles, to
the collision itself, and to a brief time after the collision, of the
order of $\tau\sim Q_s^{-1}$. Since the subsequent stages are usually
described by some form of relativistic fluid dynamics, one of the CGC
goals is to provide the value of the energy momentum tensor at some
initial time $\tau_0$. But more fundamentally, since the CGC provides
a QCD-based description of the early stages of heavy ion collisions,
it is also the framework to be used in order to justify the
applicability of hydrodynamics.

In the CGC, the energy-momentum tensor admits an expansion in powers
of the strong coupling $g^2$. However, because the color sources are
strong in the saturated regime, this expansion starts by a term in
$g^{-2}$,
\begin{equation*}
T^{\mu\nu}={\colorb\frac{Q_s^4}{g^2}}\;
\Big[c_0+c_1\,{\colord g^2}+c_2\,{\colord g^4}+\cdots\Big]\; .
\end{equation*}
The coefficients $c_{0,1,2,\cdots}$ in this expansion are themselves
infinite series in $g\rho$ (which is parametrically of order one in
the saturated regime), and therefore each order corresponds to an
infinite series of Feynman diagrams.

At leading order, one can prove that the sum of the corresponding set
of diagrams can be expressed in terms of classical solutions of the
Yang-Mills equations \cite{GelisV},
\begin{equation*}
T^{\mu\nu}_{_{\rm LO}}=
\frac{1}{4}g^{\mu\nu}\,{\colord{\cal F}^{\lambda\sigma}{\cal F}_{\lambda\sigma}}
-{\colord{\cal F}^{\mu\lambda}{\cal F}^\nu{}_\lambda}
\; ,\quad
\underbrace{\big[{\colord{\cal D}_\mu},{\colord{\cal F}^{\mu\nu}}\big]={\colorb J^\nu}}_{\rm Yang-Mills\ equation}\;,\quad
\lim_{t\to -\infty}{\colord{\cal A}^\mu(t,\vec\x)}=0\; .
\end{equation*}
These equations are non-linear and in general one cannot find analytic
solutions. However, they have been solved numerically to obtain e.g.
the initial energy density released in a collision \cite{numerical}.

\begin{figure}[htb]
\begin{center}
\resizebox*{!}{1.5cm}{\includegraphics{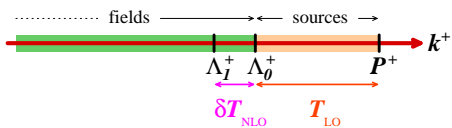}}
\end{center}
\caption{\label{fig:sep1}Slice of slow modes at NLO.}
\end{figure}

At next to leading order, the energy-momentum tensor starts being
sensitive to the slow gluon fields of the CGC effective theory, via a
loop correction. It is convenient to consider only one small slice of
these slow modes, located just below the original cutoff $\Lambda_0^+$
of the CGC, as shown in the figure \ref{fig:sep1}.
The contribution of this slice can be calculated at leading log
accuracy \cite{GelisLV},
\begin{equation*}
  {\colorc\delta T^{\mu\nu}_{_{\rm NLO}}}
  =
  \Big[
  \ln\left(\frac{\Lambda_0^+}{\Lambda_1^+}\right)\,{\colorb{\cal H}_1}
  +
  \ln\left(\frac{\Lambda_0^-}{\Lambda_1^-}\right)\,{\colorb{\cal H}_2}
  \Big]\;{\colord T^{\mu\nu}_{_{\rm LO}}}\; ,
\end{equation*}
where ${\cal H}_{1,2}$ are the JIMWLK Hamiltonians of the two
projectiles. Thanks to their form that does not mix the two
projectiles, these logarithms appear to be intrinsic properties of the
wavefunctions of the two nuclei. Therefore, they can be absorbed by
defining a new CGC effective theory that has its cutoff at the lower
values $\Lambda_1^+$,
\begin{equation*}
\Big<{\colord{\bs T}_{_{\rm LO}}}+{\colorc\delta{\bs T}_{_{\rm NLO}}}\Big>_{\Lambda_0}
=
\Big<{\colord{\bs T}_{_{\rm LO}}}\Big>_{\Lambda_1}\; ,\quad W_{\Lambda_1^\pm}\equiv \Big[1+\ln\left(\frac{\Lambda_0^\pm}{\Lambda_1^\pm}\right)\;
    {\colorb{\cal H}_{1,2}}\Big]\;W_{\Lambda_0^\pm}\; ,
\end{equation*}
provided the distribution of sources in the new effective theory is
defined as prescribed by the JIMWLK equation. This process can be
iterated, until the cutoffs become smaller than the physically
relevant scales. The outcome of this procedure is the following
formula for the energy-momentum tensor,
\begin{equation*}
\left<T^{\mu\nu}(\tau,{\colord\eta},\vec\x_\perp)\right>_{_{\rm LLog}}
=
\int 
\big[D{\colora\rho_{_1}}\,D{\colorb\rho_{_2}}\big]
\;
{\colora W_1\big[\rho_{_1}\big]}\;
{\colorb W_2\big[\rho_{_2}\big]}
\;
{T^{\mu\nu}_{_{\rm LO}}(\tau,\vec\x_\perp)}
\; ,
\end{equation*}
which resums all these logarithms. Note that at this level of
accuracy, the rapidity dependence of the left hand side comes entirely
from the JIMWLK evolution of the distributions $W$ of the two
projectiles. Besides its usefulness in heavy ion collisions, this
factorization is important because it establishes a bridge between
nucleus-nucleus collisions and other reactions such as
electron-nucleus collisions, where distributions $W$ with the same
JIMWLK evolution also appear. Let us also stress that this
factorization also applies to any sufficiently inclusive observables.
In particular, the correlation between the components of the
energy-momentum tensor measured at different points in space
\cite{GelisLV} reads
\begin{eqnarray*}
    &&
  \left<
    T^{\mu_1\nu_1}(\tau,{\colord\eta_1},\vec\x_{1\perp})
    \cdots
    T^{\mu_n\nu_n}(\tau,{\colord\eta_n},\vec\x_{n\perp})
  \right>_{_{\rm LLog}}
  =\\
  &&\qquad\quad=
  \int 
  \big[D{\colora\rho_{_1}}\,D{\colorb\rho_{_2}}\big]
  \;
  {\colora W_1\big[\rho_{_1}\big]}\;
  {\colorb W_2\big[\rho_{_2}\big]}\;
  T^{\mu_1\nu_1}_{_{\rm LO}}(\tau,\vec\x_{1\perp})
  \cdots
  T^{\mu_n\nu_n}_{_{\rm LO}}(\tau,\vec\x_{n\perp})\; .
\end{eqnarray*}
One sees that at leading log accuracy, all these correlations come
from the $W$'s (since the integrand is a product of $n$ independent
factors). This formula predicts a correlation in rapidity over a range
of order $\Delta\eta\sim \alpha_s^{-1}$, since this is the rapidity
variation necessary to produce a change in the $W$'s.

\section{Just after the collision: Glasma fields}
\label{sec:glasma}
\begin{figure}[htb]
\begin{center}
\resizebox*{!}{3.7cm}{\includegraphics{glasma_field_components.eps}}
\hfil
\resizebox*{!}{3.3cm}{\includegraphics{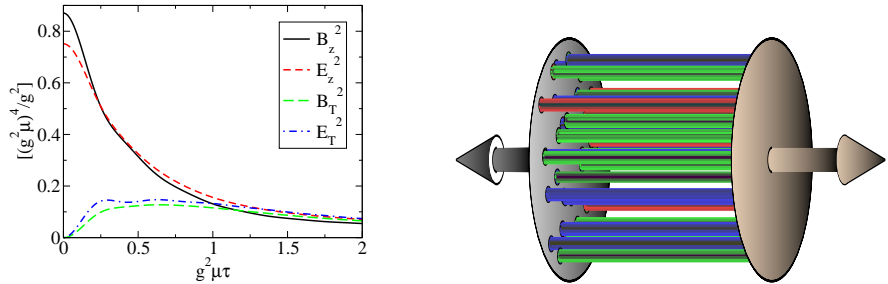}}
\end{center}
\caption{\label{fig:tubes}Left: field components evaluated by solving
  numerically the Yang-Mills equations (from \cite{LappiM1}). Right:
  longitudinal color flux tubes.}
\end{figure}
Immediately after the collision, the chromo- $\E$ and $\B$ fields have
only longitudinal components \cite{LappiM1}, forming flux tubes along
the collision axis (see the figure \ref{fig:tubes}). This
configuration of color fields has been named the {\sl glasma}. The
typical transverse size of a flux tube is of order $Q_s^{-1}$, and the
color fields are correlated over $\alpha_s^{-1}$ units of rapidity in
the longitudinal direction.

\begin{figure}[htb]
\begin{center}
\resizebox*{!}{4.5cm}{\includegraphics{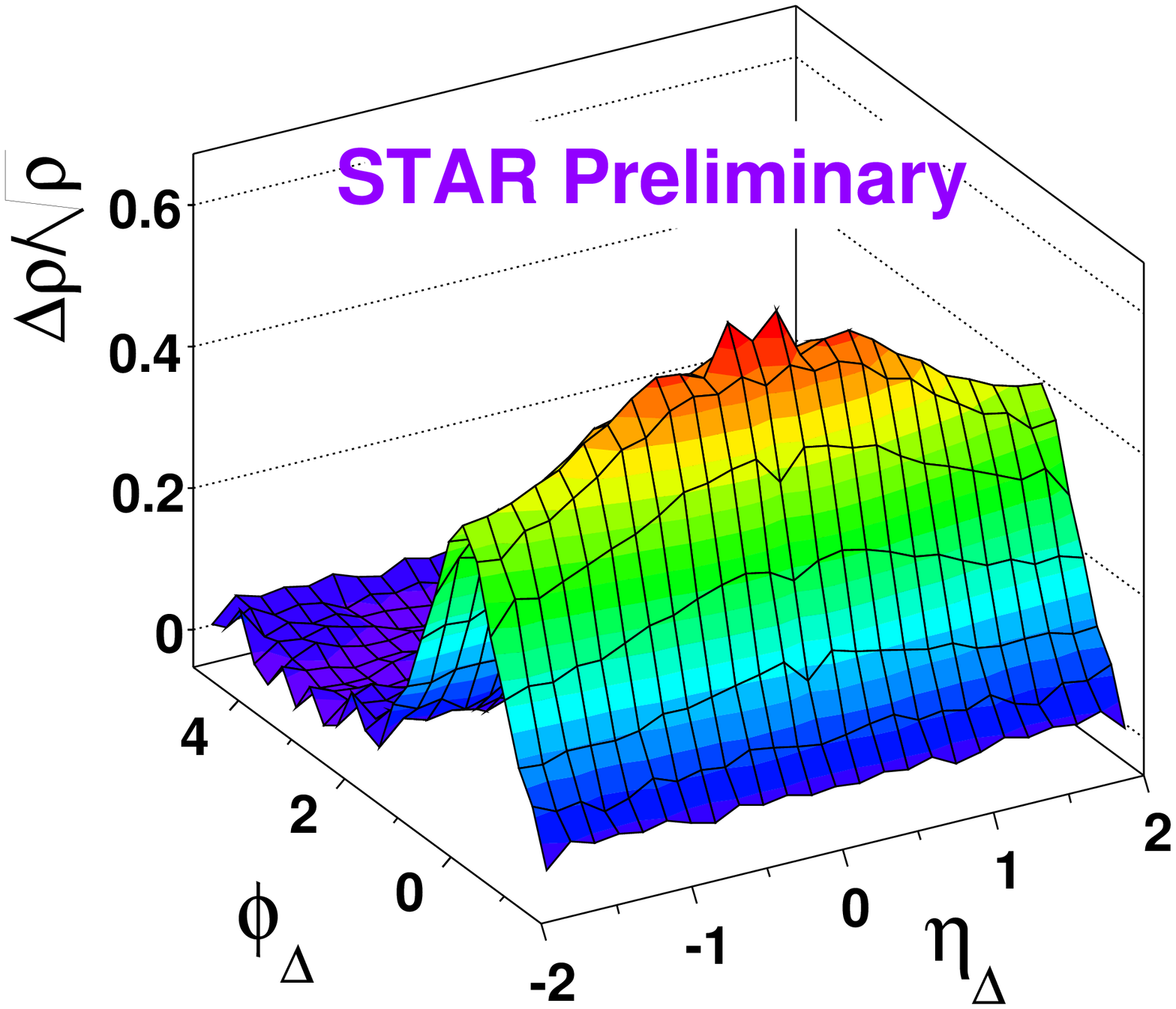}}
\hfil
\resizebox*{!}{3.5cm}{\includegraphics{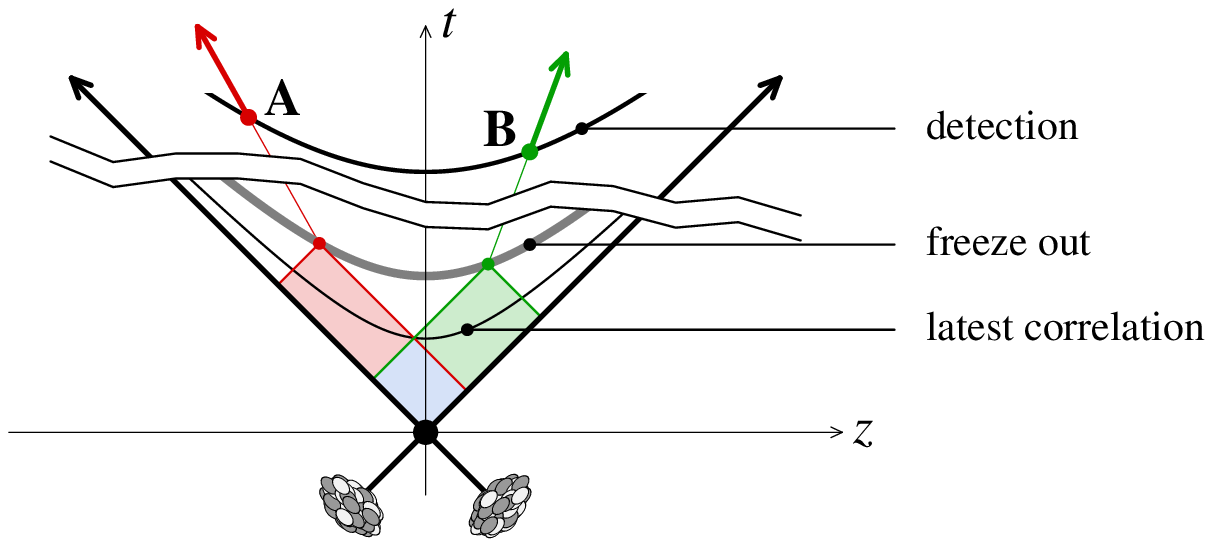}}
\end{center}
\caption{\label{fig:ridge}Left: STAR result on 2-hadron correlations. Right:
  causal relationship between two produced particles.}
\end{figure}

\begin{figure}[htb]
\begin{center}
\resizebox*{!}{3cm}{\includegraphics{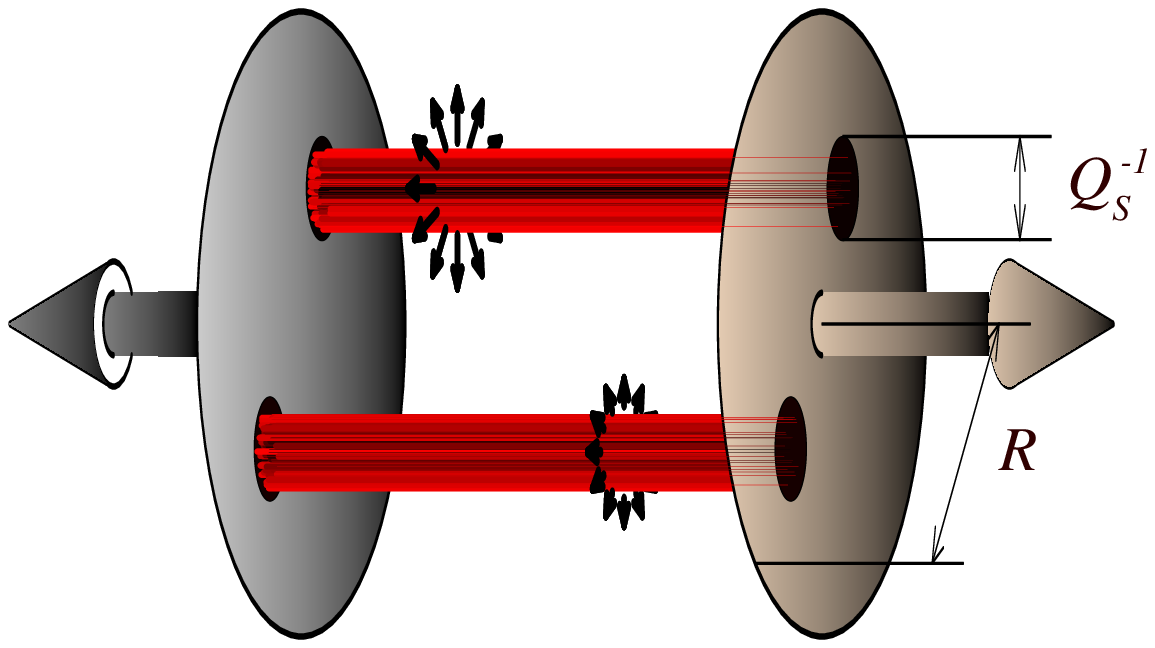}}
\hfil
\resizebox*{!}{3cm}{\includegraphics{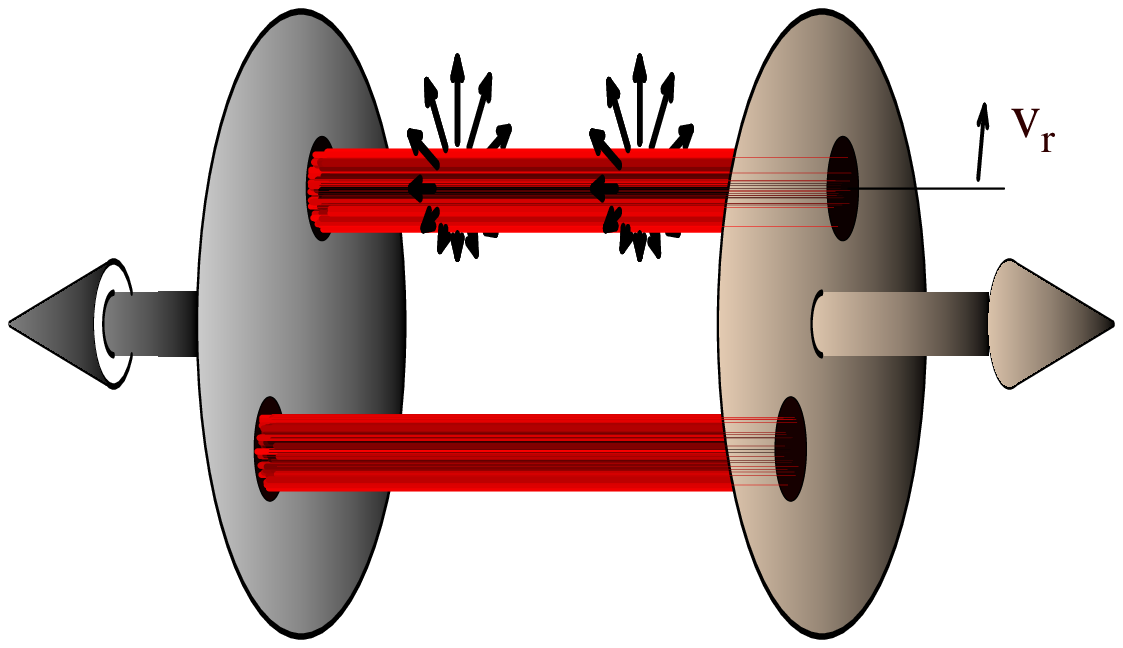}}
\end{center}
\caption{\label{fig:ridge1}Left: particle emitted from distinct tubes
  are uncorrelated. Right: collimation due to the radial flow.}
\end{figure}

This particular topology of the post-collision color fields has
several consequences, among which a peculiar form of the
energy-momentum tensor (see section \ref{sec:hydro}), the fact that
the multiplicity distribution is a negative binomial \cite{GelisLM1},
and the existence of a non-zero topological density $F\widetilde{F}$,
possibly at the origin of observable CP violating effects \cite{CP}.
But the most direct and striking consequence of these structures, when
taken as initial conditions for hydrodynamical expansion, is that they
lead to the formation of the so-called {\sl ridge} correlations, a
structure in the 2-hadron spectrum which is elongated in $\Delta\eta$
and narrow in $\Delta\phi$ (see the left plot of the figure
\ref{fig:ridge}).
By examining the causal relation between two particles separated in
rapidity (right part of the figure \ref{fig:ridge}), one can see that
the process responsible for producing a correlation between these
particles must have taken place at early times,
\begin{equation*}
    t_{\rm correlation}\le t_{\rm freeze\ out}\;\; e^{-\frac{1}{2}|\eta_{_A}-\eta_{_B}|}\; .
\end{equation*}
Since the color fields produced at early times in the CGC formalism
are correlated over rapidity intervals of order $\alpha_s^{-1}\gg 1$,
they provide a natural explanation for the rapidity dependence of the
ridge \cite{ridge}. The strength of the 2-particle correlation is
controlled by $(Q_sR)^{-2}$ --the area of one flux tube relative to
the total transverse area-- since the two particles must come from the
same tube to have been produced by the same coherent field (see the
figure \ref{fig:ridge1}, left panel).  The azimuthal dependence is
produced at a later stage, by the radial hydrodynamical flow, that
collimates the azimuthal angles of the two particles in the direction
of the radial velocity (figure \ref{fig:ridge1}, right panel).

\section{Matching to hydrodynamics}
\label{sec:hydro}
At times $\tau\gg Q_s^{-1}$, the standard description of the evolution
of the fireball is via hydrodynamical expansion. However, a trivial
consequence of the fact that the chromo- $\E$ and $\B$ fields are
initially parallel to the collision axis in the glasma is that the
energy-momentum tensor one obtains at leading order in $g^2$ in the
CGC is of the form ${\colord T^{\mu\nu}_{_{\rm LO}}(0^+,\eta,\vec\x) =
  {\rm diag}\, (\epsilon,\epsilon,\epsilon,-\epsilon) } $ where
$\epsilon$ is the energy density. The fact that the longitudinal
pressure is negative is problematic for a smooth matching to
hydrodynamics, because it means that the CGC energy-momentum tensor is
quite far from the form for which quasi-ideal hydrodynamics is
applicable (i.e. one where the spatial part of the tensor is not too
far from being proportional to the identity).

It has been known for some time that corrections to the leading order
CGC prediction suffers from secular divergences. Indeed, the solutions
of classical Yang-Mills equations are usually unstable, as noticed in
several works \cite{insta}. This implies that the next to leading
order correction to the energy momentum tensor calculated in the CGC
framework grows with time, and eventually becomes larger than the
leading order. It is possible to resum a subset of these higher order
corrections, that at each loop order picks up the most singular of
these contributions. This resummation amount to superimposing a
fluctuation to the initial condition at $\tau=0^+$ for the classical
glasma field \cite{resum},
\begin{equation*}
T^{\mu\nu}_{\rm resummed}(\tau,\eta,\vec\x_\perp)
=
\int\big[D{\colora\rho_{_1}}\,D{\colorb\rho_{_2}}\big]\big[D a\big]\;
{\colord F[a]}\;
{\colora W_{1}[\rho_{_1}]}\,{\colorb W_{2}[\rho_{_2}]}\;
T^{\mu\nu}_{_{\rm LO}}[\underbrace{{\cal A}+{\colord a}}_{\rm initial\ field}]
\end{equation*}
with a Gaussian distribution $F[a]$ for this fluctuation. The effect
of this resummation has not been fully investigated in the context of
the CGC due to some technical difficulties that preclude a
straightforward numerical evaluation of the previous formula. However,
it is possible to devise a scalar toy model where the effect of such
fluctuations can be studied.
\begin{figure}[htb]
\begin{center}
\resizebox*{!}{4cm}{\rotatebox{-90}{\includegraphics{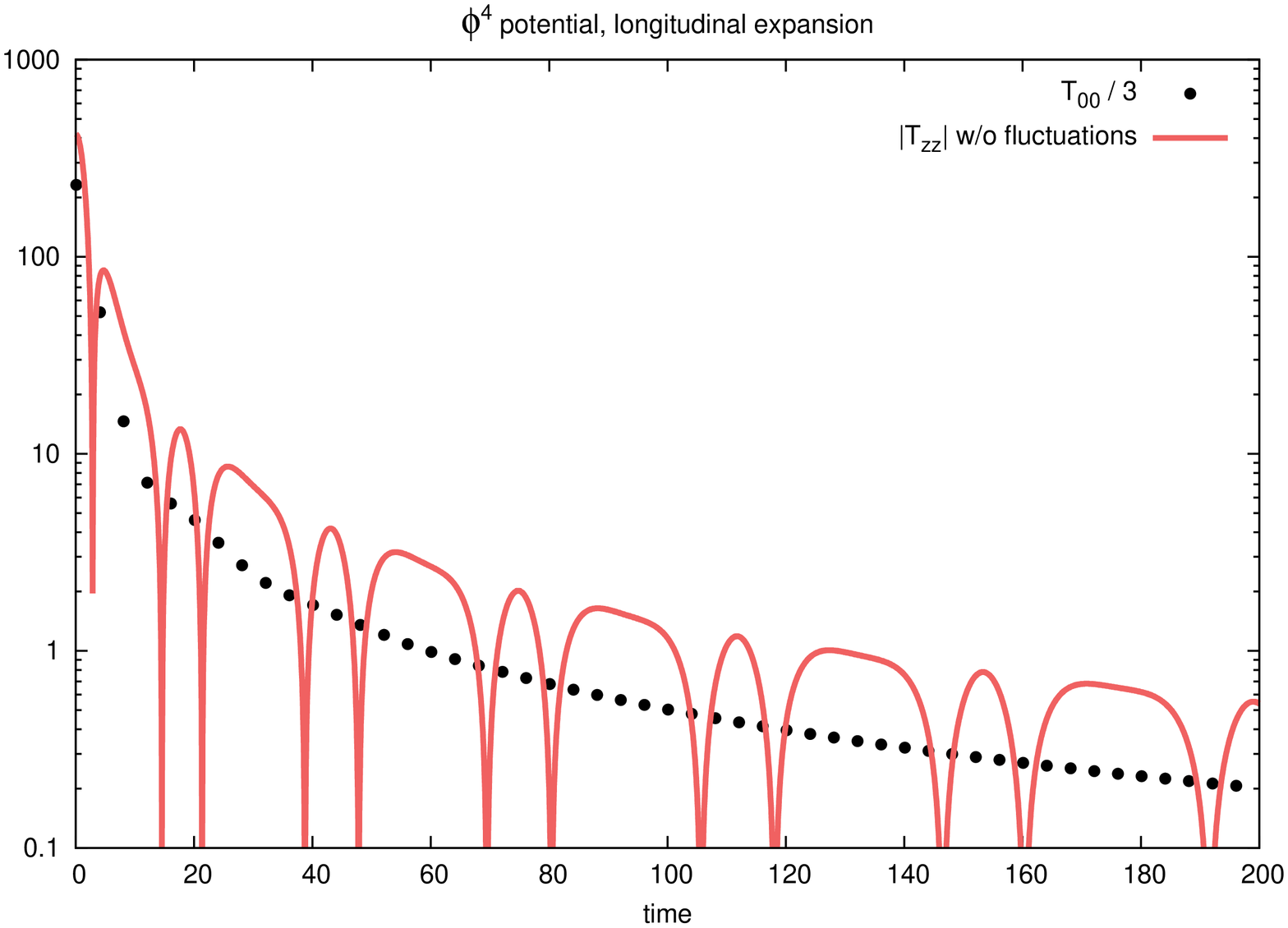}}}
\hfil
\resizebox*{!}{4cm}{\rotatebox{-90}{\includegraphics{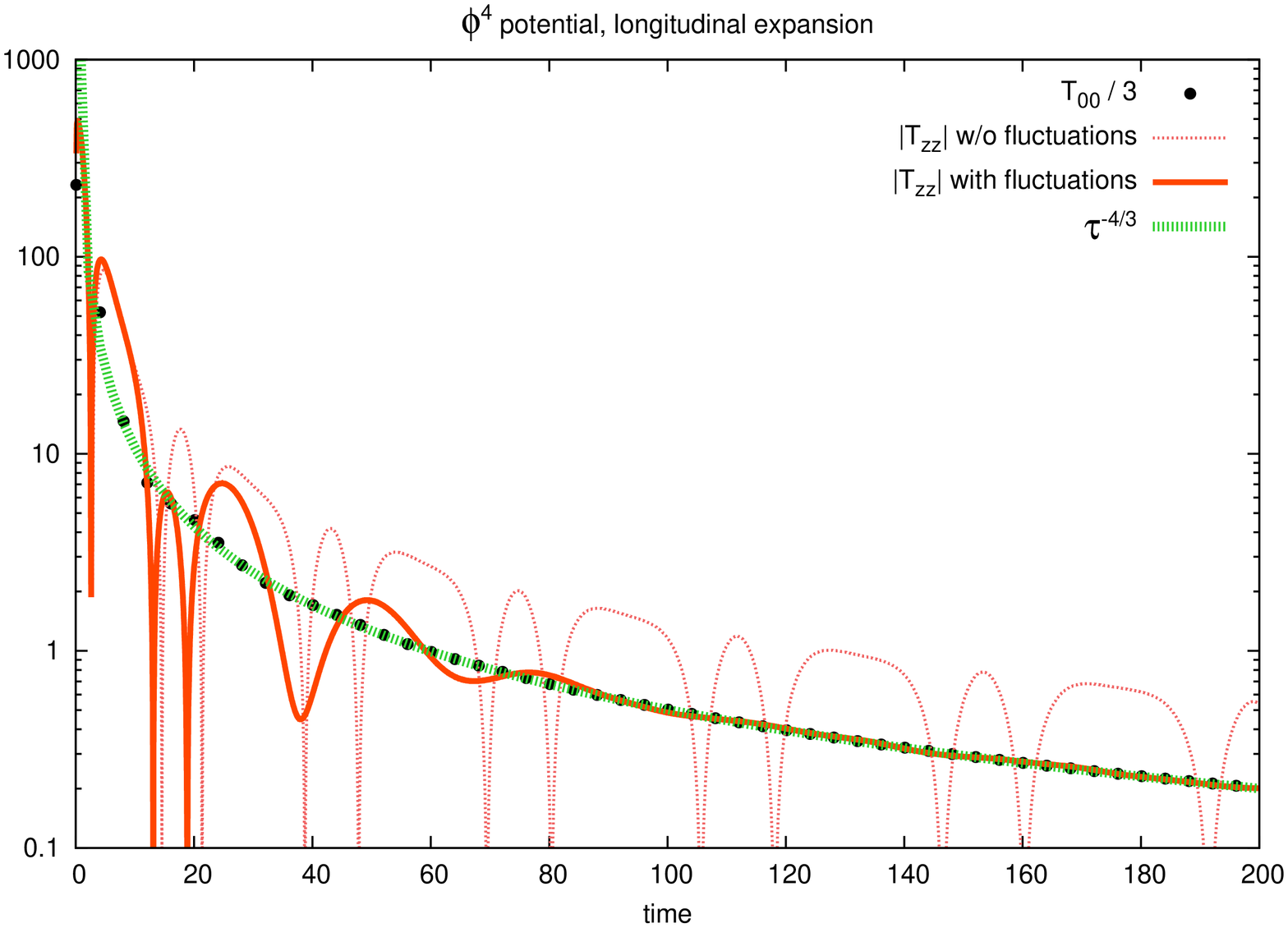}}}
\end{center}
\caption{\label{fig:fluct}Left: evolution of $\epsilon$ and $p$ without fluctuations of the initial field. Right: evolution with initial fluctuations.}
\end{figure}
To keep things extremely simple
in this toy model, we discard any spatial dependence, so that the
classical equation of motion reads
\begin{equation*}
\ddot\phi+\frac{1}{\tau}\dot\phi-{\bs\nabla}_\perp^2\phi+V^\prime(\phi)=0\; ,
\end{equation*}
and we take a quartic potential $V(\phi)\sim\phi^4$. The term
$\dot\phi/\tau$ is due to longitudinal expansion in the rapidity
direction, as would be the case in a collision process. The time
evolution is started at some some initial proper time $\tau=\tau_0$,
and we average the energy-momentum tensor over Gaussian fluctuations
of the initial field.  We first did this toy calculation without
fluctuations of the initial field (see the left plot of the figure
\ref{fig:fluct}) and there one sees that the pressure oscillates with
time: there is no single-valued relationship between the energy
density and the pressure, which means that there is no equation of
state in this case. In the right plot of the figure \ref{fig:fluct},
we show the effect of the initial field fluctuations: the oscillations
of the pressure are now damped, and it relaxes towards $\epsilon/3$ --
which is the equation of state expected for a scale invariant theory
in four dimensions. Moreover, one also observes that the energy
density decreases as $\tau^{-4/3}$, which is the expected behavior in
hydrodynamics if the longitudinal pressure is $\epsilon/3$. Thus, in
this toy model, it appears that the fluctuations of the initial field
play a crucial role in the relaxation of the system towards
hydrodynamical behavior.

\section*{Conclusions}
The Color Glass Condensate provides a first principles framework to
study the early stages of high energy nuclear collisions, that involve
gluons in the saturated regime. It correctly describes the energy
density released in such a collision, as well as important features of
the correlations observed in the final state. In this context, the
most pressing and challenging question for future work is arguably
that of thermalization.
\vskip 3mm
\noindent{\bf Acknowledgements:} This work is supported in part by Agence
Nationale de la Recherche via the programme ANR-06-BLAN-0285-01.

\end{document}